\documentclass[useAMS,usenatbib]{mn2e}

\usepackage{graphicx}

\title[Filament eruption with apparent reshuffle of endpoints]
{Filament eruption with apparent reshuffle of endpoints}
\author[Boris Filippov]{Boris Filippov \thanks{E-mail:
bfilip@izmiran.ru} \\ Pushkov Institute of Terrestrial Magnetism,
Ionosphere and Radio Wave Propagation of the Russian Academy of
Sciences (IZMIRAN), \\ Troitsk, Moscow 142190, Russia}

\begin{document}

\date{Accepted 0000 December 15. Received 0000 December 14; in original form 0000 October 11}

\pagerange{\pageref{firstpage}--\pageref{lastpage}} \pubyear{2002}

\maketitle

\label{firstpage}

\begin{abstract}
Filament eruption on 30 April - 1 May 2010, which shows the
reconnection of one filament leg with a region far away from its
initial position, is analyzed. Observations from three viewpoints
are used for as precise as possible measurements of endpoint
coordinates. The northern leg of the erupting prominence loop
'jumps' laterally to the latitude lower than the latitude of the
originally southern endpoint. Thus, the endpoints reshuffled their
positions in the limb view. Although this behaviour  could be
interpreted as the asymmetric zipping-like eruption, it does not
look very likely. It seems more likely  to be reconnection of the
flux-rope field lines in its northern  leg with ambient coronal
magnetic field lines rooted in a quiet region far from the
filament. From calculations of coronal potential magnetic field,
we found that the filament before the eruption was stable for
vertical displacements, but was liable to violation of the
horizontal equilibrium. This is unusual initiation of an eruption
with combination of initial horizontal and vertical flux-rope
displacements showing a new unexpected possibility for the start
of an eruptive event.
\end{abstract}


\section{Introduction}

Many solar filaments (or prominences when they are observed above
the solar limb) end their life with a sudden rapid rise called an
eruption. Sometimes a filament rises like an enlarging loop lying
in a plane containing filament endpoints anchored in the
chromosphere and the centre of the Sun \citep{b16}. A famous
example suggestive of such behaviour is the eruption of the giant
prominence on 28 June 1945 ('Granddaddy') observed at the High
Altitude Observatory. Some eruptive prominences deviate
significantly from this plane and move in a non-radial direction
\citep{b17,b11, b12, b31}. Moreover, the loop may be not flat but
the apex exhibits writhing motion as it rotates about the
direction of ascent \citep{b21,b29, b36, b38, b19, b27, b23}.

There are also partial filament eruptions when only a section of a
long filament starts to ascend while other parts of the filament
are observed unchanged \citep{b33}. Usually a prominence stretched
along the limb consists of a number of arches with feet (called
also barbs) connected to the chromosphere like a long road bridge
with several spans. After the start of ascending, the feet
successively break one after another except the filament endpoints
for a full eruption or feet of undisturbed filament sections for a
partial eruption. Of course, in real solar environment, events are
often asymmetric. One leg of an eruptive prominence may be fixed
to the chromosphere at the prominence endpoint while the other
point connecting the prominence with the chromosphere changes its
position following after successive breaking of intermediate feet.
\citet{b25} identified two types of asymmetric filament eruptions:
whipping-like, where the active leg whips upward, occasionally
extending high into the corona; and zipping-like, where the
visible end of the active leg moves along the polarity inversion
line (PIL) like the unfastening of a zipper. It should be noted
that the visibility of a filament depends on the loading of its
magnetic skeleton with dense plasma. During the asymmetric
filament eruption, the active leg can either whip upward, if it is
anchored at the location where the eruption initiates; or 'zip'
away from the visible end of the active leg, where the eruption
initiates, toward the 'invisible' end of the active leg. The
'invisible' end later becomes visible during the zipping process
with mass draining down along axial filament field lines.

\begin{figure}
\includegraphics[width=84mm]{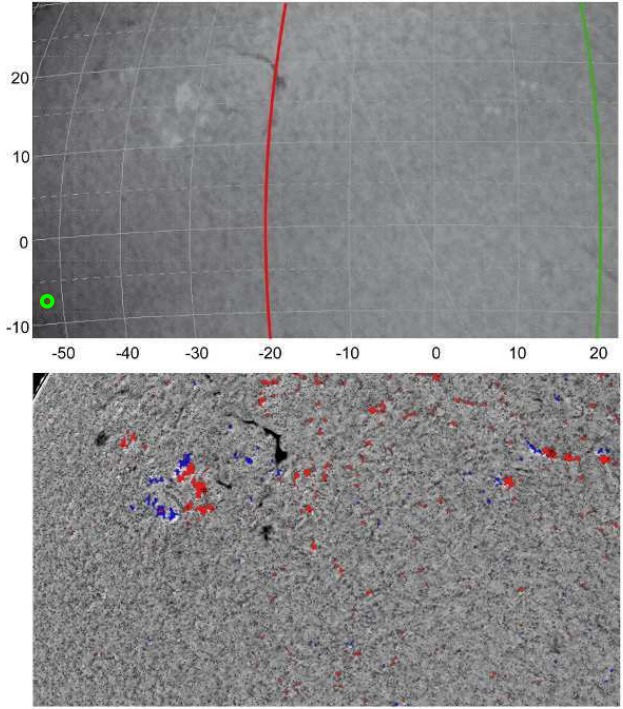}
\caption{Position of the filament on the solar disk before the
eruption. Top: H$\alpha$ image at 16:15 UT (courtesy of the Meudon
Observatory). The red line shows the position of the east limb for
{\it STEREO A}, the green line indicates the position of the west
limb for {\it STEREO B}. The green circle shows the final position
of the new filament end region before the filament faded. Bottom:
H$\alpha$ image at 16:34 UT (courtesy of the Kanzelhoehe Solar
Observatory) overlaid with line-of-sight field strength from
\textsl{SOHO}/MDI (only field strengths $> 50$~Gauss are shown;
red -- negative; blue -- positive (courtesy of the {\it SOHO}/MDI
consortium)). }
\end{figure}

While, falling back to the chromosphere, eruptive prominence
material can move along pre-existing magnetic flux tubes, possibly
highly stretched and deformed, which belong to the prominence
magnetic skeleton, there are also indications that sometimes
prominence plasma returns to the chromosphere along trajectories
that were formed by reconnection of the prominence magnetic field
with the ambient coronal magnetic field. \citet{b18} reported an
explosive filament eruption on 13 July 2004 in which one part of
the ejecta escaped as a CME, and another one fell back onto the
Sun. The latter part consisted of fragments of the filament
dispersed into a cloud covering almost the whole NW quadrant of
the solar disk. Obviously, some areas where filament fragments
landed had not been connected by field lines with the filament. On
7 June 2011, an active region filament near the west solar limb
rose and erupted, hurling an enormous amount of material into the
solar atmosphere \citep{b20, b15, b5, b34} . The diagonal scale of
the ejecta appears at least an order of magnitude larger than the
initial foot-point separation, and suggests the filament carried a
very large amount of mass. A significant fraction of the
prominence mass was observed falling back to the solar surface
along newly reconfigured magnetic field lines. \citet{b34}
consider this event as clear evidence that large-scale
re-configuration of the coronal magnetic field takes place during
solar eruptions via the process of magnetic reconnection.
\citet{b25c} reported a disk event on 25 October 1994, which
provided evidence for a large-scale magnetic reconnection
occurring between the expanding twisted loops and overlying
transequatorial loops that interconnect quiet solar regions.

In this work, we analyze observations of the eruptive filament on
30 April - 1 May 2010, which shows reconnection of one filament
leg with a region far away from its initial position. In contrast
to above-mentioned examples of the coronal reconnection, the
erupting filament loop does not disintegrate but keeps its shape
of a rather thin loop even after a fast 'jump' of the endpoint
over a distance of 0.7 solar radius $R_\odot$. We calculated
parameters of coronal potential magnetic field and found that the
eruption began with instability not in the vertical direction, as
typical for eruptive filaments, but after violation of the
horizontal equilibrium.

\section[]{Observations of the Eruptive Filament on 30 April - 1 May 2010 from Three Viewpoints}

A quiescent filament, located close to NOAA active region 11064 to
the north-west from it (Fig. 1), started to rise rapidly after 23
UT on 30 April 2010. The eruption was observed on the disk in
H$\alpha$ line at the Mauna Loa Solar Observatory with a cadence
of 3 minutes (Fig. 2) and Culgoora Solar Observatory with a
cadence of 1 minute as well as on the eastern limb by the {\it
Solar Terrestrial Relations Observatory - Ahead (STEREO A)} and
close to the western limb by {\it STEREO B (Behind)} with a
cadence of 10 minutes (Fig. 3 and Fig. 4 ). The event was also
observed on the disk with the {\it PROBA2}/SWAP EUV solar
telescope in a spectral bandpass centered on 174  \AA\  with a
cadence of about 1 minutes \citep{b29aa,b19a}. At first, it looked
like a typical eruption of a filament showing an expanding loop
with anchored endpoints (movie 1). At 23:46 UT, the top of the
loop folded over in images obtained by Sun Earth Connection
Coronal and Heliospheric Investigation (SECCHI) EUVI \citep{b37,
b14} onboard {\it STEREO A}, showing writhing of the filament
axis. The apex of the eruptive filament (prominence) deflected to
the South during the ascending. At 00:06 UT, it was over the
southern endpoint of the prominence. This endpoint became wider
and consisting of several strands of threads after 00 UT. The
southernmost strand faded out after 00:30 UT, while the northern
strand became narrower with fine threads.

\begin{figure}
\includegraphics[width=84mm]{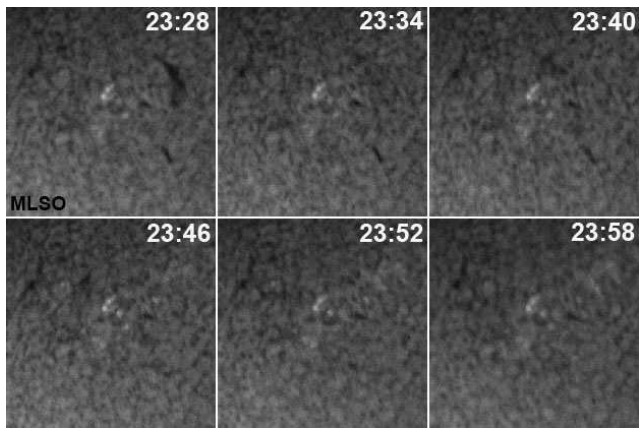}
\caption{Filament eruption in H$\alpha$ line. The size of each
frame is .5 $R_\odot$ $\times$ .5 $R_\odot$. (Courtesy of the
Mauna Loa Solar Observatory, operated by the High Altitude
Observatory, as part of the National Center for Atmospheric
Research (NCAR). NCAR is supported by the National Science
Foundation.)}
\end{figure}

The behaviour of the northern endpoint of the filament was more
dramatic. Before the eruption, the filament axis entered the
chromosphere at this endpoint inclined slightly to the South from
the vertical. From 23:30 UT to 00 UT, the axis was nearly
vertical. Then, it inclined to the South. About 00:30 UT, the
northern leg of the erupting prominence loop 'jumped' laterally to
the position further south than the southern endpoint of the
prominence. The loop proceeded with expanding and ascending,
however the endpoints reshuffled their positions in {\it STEREO A}
images. The originally southern endpoint is now on the North,
while the originally northern endpoint is on the South. Further
evolution of the eruptive prominence in {\it STEREO A} images
looks like typical prominence eruption with anchored in the
chromosphere endpoints, if one forgets that the endpoints have
exchanged the roles. The top of the prominence leaved the {\it
STEREO A} field-of-view at 00:40 UT, and after 01:40 UT all the
structure faded and flew away. After half an hour acceleration
($\sim$ 23 UT - 23:30 UT), the top of the prominence rose with an
approximately constant speed of 90 km s$^{-1}$ (Fig. 5).

\begin{figure}
\includegraphics[width=84mm]{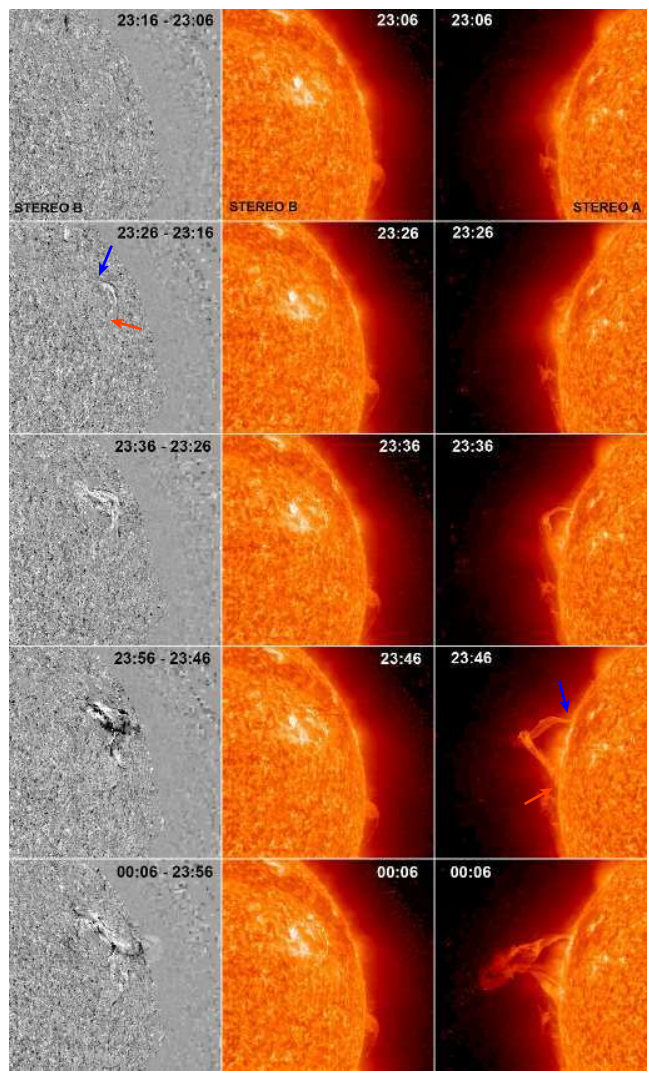}
\caption{Filament eruption observed on 30 April - 1 May 2010 by
{\it STEREO A} (right column)  and {\it STEREO B} (middle column)
in the 304 \AA \ channel. The left column represents {\it STEREO
B} 304 \AA \ running-difference images. The blue (red) arrow
points to the position of the originally northern (southern)
filament endpoint. The size of each frame is 1.1 $R_\odot$
$\times$ 1.1 $R_\odot$. (Courtesy of the {\it STEREO}/SECCHI
Consortium, NASA/SDO and the AIA science team.)}
\end{figure}

\begin{figure}
\includegraphics[width=84mm]{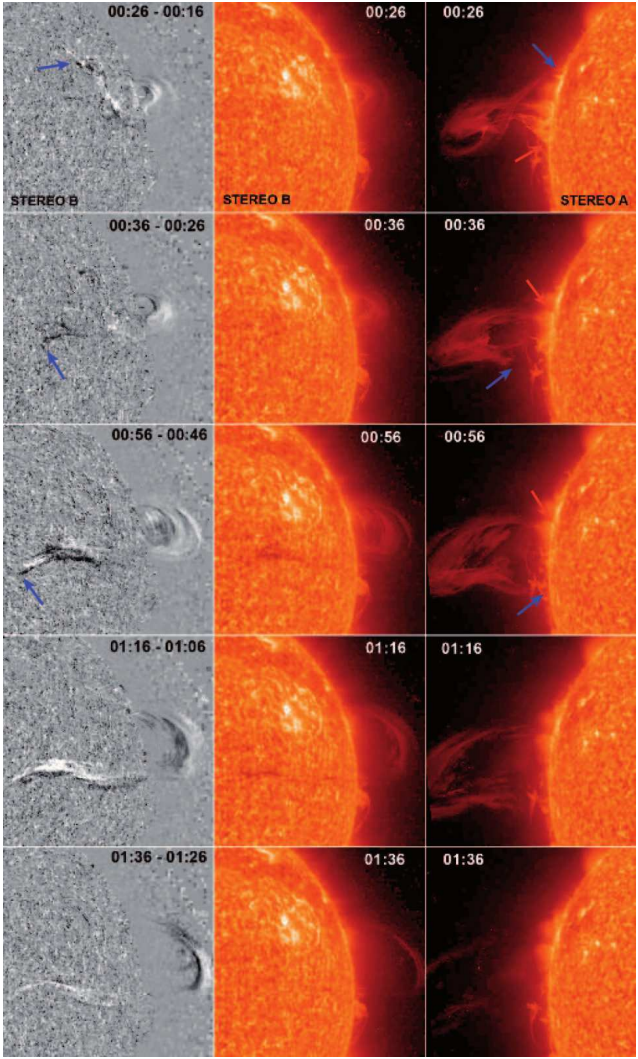}
\caption{Continuation of the sequence shown in  Fig. 3.}
\end{figure}

The eruption of the prominence was followed by a coronal mass
ejection (CME), which was observed from all three viewpoints (Fig.
6). Due to the geometrical factor, it was brightest in the
field-of-view of the {\it STEREO A} COR2 coronagraph (Howard et
al., 2008). At the late stage, the legs of the CME connecting the
core with the Sun show noticeable twisted structure (not visible
in the difference image in Fig. 6). The Large Angle and
Spectrometer Coronagraph (LASCO) C2  \citep{b4}  onboard the {\it
Solar and Heliospheric Observatory (SOHO)} and the {\it STEREO B}
COR2 coronagraph registered only faint features, which became
perceptible enough only in difference images. The CME propagated
in the SEE direction for {\it STEREO A} and SOHO, while in the SWW
direction for {\it STEREO B}. The bright core in the {\it STEREO
A} COR2 field-of-view moved with a speed of about 100 km s$^{-1}$.
The frontal CME structure, of course, moved faster. According to
the {\it SOHO}/LASCO CME Catalog
(http://cdaw.gsfc.nasa.gov/CME\_list/), the CME appeared first in
the field-of-view of C2 at 07:12 UT on 1 May at a polar angle of
111$^\circ$, had an angular width of 108$^\circ$, and reached the
final speed of 380 km s$^{-1}$.

\begin{figure}
\includegraphics[width=84mm]{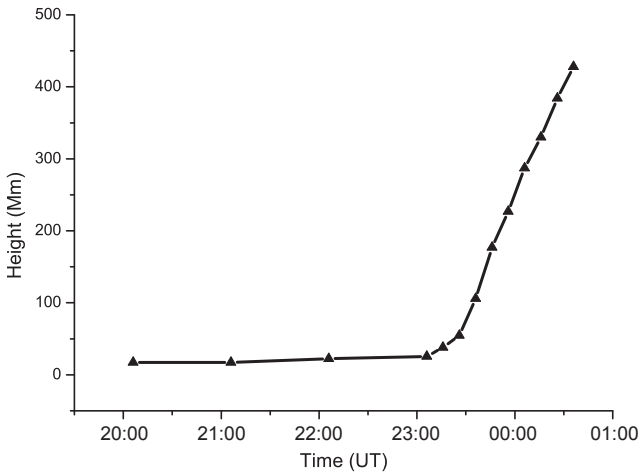}
\caption{Filament top height above the {\it STEREO A} eastern limb
as a function of time.}
\end{figure}

\begin{figure}
\includegraphics[width=84mm]{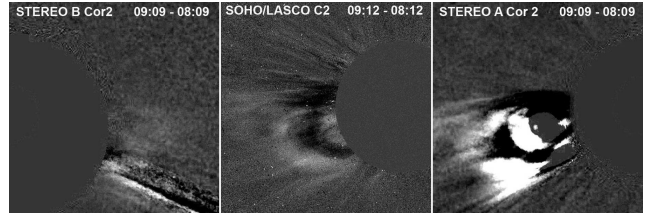}
\caption{Difference images of the coronal mass ejection associated
with the filament eruption in the field-of-view of three
coronagraphs. (Courtesy of the {\it STEREO}/SECCHI Consortium and
the {\it SOHO}/LASCO Consortium, ESA and NASA.)}
\end{figure}

In H$\alpha$ images obtained with groundbased telescopes, the
filament rotates counterclockwise about its southern end point
like a nearly straight structure (Fig. 2, see also movie 2). It
passed over AR 11064 in projection and faded when it stretched in
nearly longitudinal direction. Two short faint flaring ribbons
appeared at 23:50 UT at the location of the most curved and widest
section (an elbow in Fig. 1) of the pre-eruptive filament (Fig.
2).

Brightening below the ascending filament in {\it STEREO B} images
arose at 23:36 UT soon after the start of the eruption  (Fig. 3,
Fig. 4, movie 3). Later on, two discontinuous ribbons became very
prominent in images and a system of post-eruptive loops was
observed in 171 \AA \ channels of both {\it STEREO A} and {\it
STEREO B} spacecraft. The beginning of the event in {\it STEREO B}
observations also looks like a typical filament eruption. Between
00:26 UT and 00:36 UT, the attachment of the filament to the
northern endpoint tore, the filament end jumped a large distance
both in latitude and longitude, and found new more or less stable
position in projection on the disk. In order to understand, what
happened with the connectivity of the filament, what was a real
displacement, and what was a projection effect, let us examine in
details the position of the filament endpoints.

\section[]{Filament Endpoint Coordinates}

We used images in JPEG format from  websites of the {\it STEREO}
mission and ground observatories. All images are rotated in such a
way that the projection of the solar rotation axis is vertical and
the heliographic North is on the top of the images. In the
vertical plane containing the line of sight, the rotation axis is
inclined by an angle $B_0$ from the sky plane. Usually $B_0$ is
indicated as the heliolatitude of the centre of the solar disk in
an image of the Sun. We will use the Heliocentric Earth Equatorial
coordinate system (HEEQ) with $Oz$ directed along the solar
rotation axis, $Ox$ pointed to the intersection of solar equator
and solar central meridian as seen from Earth, and $Oy$ completing
the right-handed system. After measurements of the coordinates
$y'$ (horizontal) and $z'$ (vertical) relative the to disk centre
in units of the solar radius $R_\odot$ (since the scale is
different in different observational data), we should rotate the
$Ox'y'z'$ coordinate system by the angle $B_0$ around the $Oy'$
axis and transform the Cartesian coordinates $x, y, z$ into
spherical heliocentric coordinates $\varphi$ and $\lambda$
assuming that all points of interest are located on the spherical
surface $r = R_\odot$.

\begin{equation}
x = x^\prime \cos B_0 - z^\prime \sin B_0,
\end{equation}

\begin{equation}
y = y^\prime,
\end{equation}

\begin{equation}
z = x^\prime \sin B_0 + z^\prime \cos B_0,
\end{equation}

\begin{equation}
\sin \varphi = \frac{z}{R_\odot} = \frac{z^\prime}{R_\odot} \cos
B_0 + \sin B_0 \sqrt{1 - \frac{y^\prime + z^\prime}{R_\odot^2}},
\end{equation}

\begin{equation}
\sin \lambda = \frac{y^\prime}{R_\odot \cos \varphi},
\end{equation}

The results of measurements are shown in Fig. 7 and Fig. 8. The
red curves and symbols correspond to the southern endpoint, the
blue curves and symbols correspond to the northern endpoint.
Triangles (squares) show data from {\it STEREO A (STEREO B)}
images, circles represent data from H$\alpha$ observations. In
general, data from all points of view are in good agreement with
each other. Uncertainties and discrepancies appear due to
difficulties of identification of the same features in different
projections because of the complicated internal structure of the
filament. In particular, bifurcation of the southern filament leg
leads to appearing of an additional (higher-latitude) branch in
the southern endpoint position data of {\it STEREO A} (Fig. 7). On
the late phase of the eruption, the originally southern endpoint
cannot be recognized in  {\it STEREO-B} images. Only material
projected on the sky plane above the limb as a prominence
indicates the position of the originally southern section of the
filament. The lowest part of the prominence, indicated by the
magenta arrow in Fig. 9, is the nearest to the southern endpoint
visible section of the filament. The magenta curve in Fig. 7 shows
the latitude of the crossing point of the prominence leg with the
{\it STEREO-B} limb. Due to the curved shape of the erupting
filament, the latitude of the crossing point is greater than the
latitude of the southern endpoint derived from {\it STEREO A}
observations (short red curve with triangle symbols in Fig. 7) but
the difference is not very significant. There is a wide jump of
more than 30$^\circ$ in the northern endpoint latitude during
changing of its 'connectivity'.

\begin{figure}
\includegraphics[width=84mm]{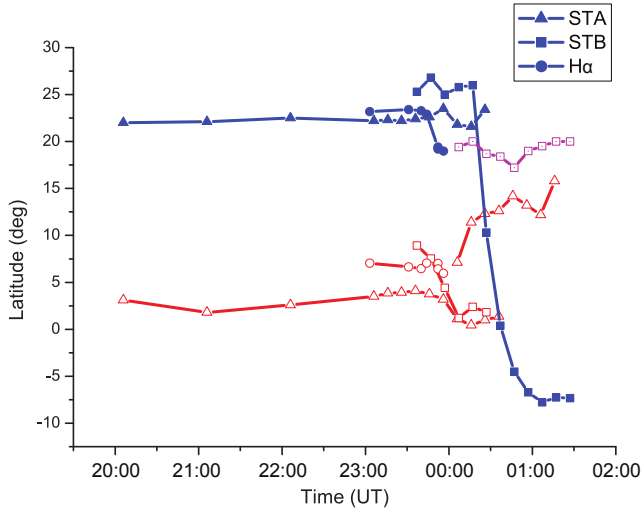}
\caption{Filament endpoint latitude derived from observations from
different points of view. The red lines and symbols correspond to
the southern endpoint, the blue lines and symbols correspond to
the northern endpoint. The magenta curve  shows the latitude of
the crossing point of the prominence leg with the  {\it STEREO-B}
limb.}
\end{figure}

\begin{figure}
\includegraphics[width=84mm]{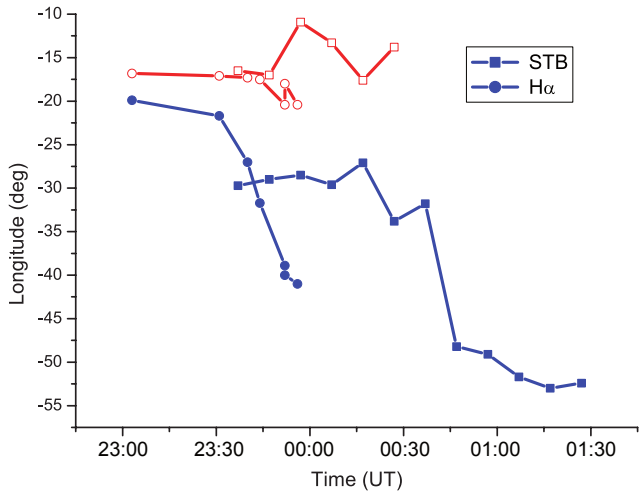}
\caption{Filament endpoint longitude derived from observations
from different points of view. The red lines and symbols
correspond to the southern endpoint, the blue lines and symbols
correspond to the northern endpoint.}
\end{figure}

Longitude of the northern endpoint also changes about 30$^\circ$
(Fig. 8). The exact position of this endpoint in longitude is less
definite than in latitude because the northern section of the
filament has nearly longitudinal orientation (Fig. 1), while this
section is rather faint in H$\alpha$ images and its visibility
varies in time (compare Fig. 1 and Fig. 2). This is the main
reason for a more steep decrease of the northern endpoint
longitude derived from the H$\alpha$ data. We fix the visible
northern end of the filament in H$\alpha$ images, however this is
not the endpoint of the flux rope anchored in the photosphere but
the middle part of the moving flux rope with the northern endpoint
still anchored at the same place (compare Fig. 2 or better movie 2
with the difference images at 23:56 UT - 00:06 UT, 00:26 UT -
00:16 UT in Fig. 3, Fig. 4 and movie 3).

Formulas (4) - (5) and results shown in Fig. 7 and Fig. 8 assume
that selected points are located on the solar surface. This is
correct for the southern endpoint which is observed connected with
the chromosphere  on the limb till the last phase of the event.
Situation with the northern endpoint is more intricate. We have no
information about the height of the filament loop end above the
chromosphere observed by {\it STEREO B} . Since the final position
of the moving loop end is not far from the disk centre in the {\it
STEREO B} images, the real coordinates more or less correspond to
the values shown in Fig. 7 and Fig. 8. We can only estimate an
upper bound on the height of the filament end point seen by {\it
STEREO B}. The difference {\it STEREO A} image in Fig. 9 shows
that at 01:06~UT the former northern (now eastern) end point must
have a height lower than the line of sight tangent to the limb for
\textsl{STEREO A} because the prominence material extends down to
the limb in the \textsl{STEREO A} view (see also movie 3). At this
time, the eastern (former northern) endpoint in {\it STEREO B}
images finds its final position.

\begin{figure}
\includegraphics[width=84mm]{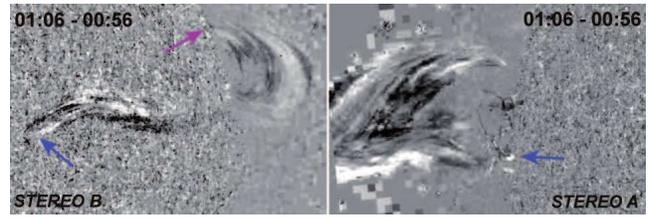}
\caption{Difference images of the eruptive filament at 01:06 UT-
00:56 UT. The blue arrows show the position of the observed for
the last time originally northern endpoint. The magenta arrow
shows the crossing point of the prominence leg with the  {\it
STEREO-B} limb. The size of each frame is 1 $R_\odot$ $\times$ .7
$R_\odot$.}
\end{figure}

The height of the line-of-sight above a point on the solar surface
with coordinates $\varphi$ and $\lambda$ (or the length of the
radial segment from a given point to the intersection with a
tangent to the limb) is
\begin{equation}
h = R_\odot (\frac{1}{ \cos \gamma} -1),
\end{equation}
where $\gamma$ is the angle between the radius passing through the
given point and the sky plane. This angle is given by

\begin{equation}
\cos \gamma = \sqrt{\sin^2 \varphi + \cos^2 \varphi \sin^2
\lambda}.
\end{equation}

For {\it STEREO A} images, we should use a coordinate system
related to them. For this purpose, the HEEQ system should be
rotated by the separation angle between the Earth and {\it STEREO
A} of 70$^\circ$ around the $Oz$ axis and by the angle of
inclination of the solar rotational axis from the sky plane $B_0$
= 4$^\circ$.12 around the $Oy$ axis. According equations (1) - (5)

\begin{equation}
\sin \varphi_A = \sin \varphi \cos B_0 + \cos \varphi \cos
(\lambda + \Delta \lambda) \sin B_0,
\end{equation}

\begin{equation}
\sin \lambda_A = \frac{\cos \varphi}{\cos \varphi_A} \sin (\lambda
+ \Delta \lambda) .
\end{equation}

The approximate coordinates of the eastern endpoint $\varphi$ =
8$^\circ$, $\lambda$ = -52$^\circ$ in the HEEQ system transform to
$\varphi_A$ = 10$^\circ$, $\lambda_A$ = -121$^\circ$. Then,
$\gamma_A$ = 31$^\circ$ and $h$ = 116 Mm. This height is much less
than the height of the prominence top of about 500 Mm  at this
time (Fig. 5). Therefore, the shape of the prominence axis at the
later stage of the eruption is a loop with low endpoints. The
eastern (former northern) filament extremity is located definitely
below $\sim$ 100 Mm and possibly finds connection with the
chromosphere at a new point after coronal reconnection.

\section[]{Magnetic Field and Filament Stability}

It is widely accepted now that filaments represent cold dense
plasma contained within magnetic flux ropes embedded into coronal
magnetic field. A flux rope can exist in equilibrium in an ambient
magnetic field for a rather long time before an eruption.
\citet{b35}  showed first that there is a critical height for
stable flux rope equilibria above which the background coronal
magnetic field decreases faster than the inverse height. The
transition from stability to instability was named later
catastrophic loss of equilibrium and was assumed to be the cause
of sudden eruptive events \citep{b28, b13, b24, b30}. \citet{b35}
modeled a flux rope with the magnetic field created by a straight
line current. If a flux rope is curved, an additional force called
the 'Lorentz self-force' or 'hoop force' is present \citep{b3} .
It is directed away from the curvature centre. In the presence of
an ambient magnetic field, the curved flux rope can be both in
stable or unstable equilibrium depending on properties of the
external field. \citet{b22}  called the related instability 'torus
instability' and showed, following \citet{b3} , that it occurs
when the background magnetic field decreases along the major
radius $R$ of the expanding flux rope faster than $R^{-1.5}$.
\citet{b8} carefully compared the two types of models and came to
conclusion that the same physics is involved in the instabilities
of circular and straight current channels. The stability of the
flux-rope equilibrium in both models depends on the rate of the
background field decrease, quantified by the so-called decay
index,

\begin{equation}
n = - \frac{\partial \ln B_t}{\partial \ln h},
\end{equation}
where $B_t$ is the horizontal magnetic-field component
perpendicular to the flux rope axis and $h$ is the height above
the photosphere. \citet{b9a,b10} pioneered in applying this index
to the analysis of filament stability. \citet{b8} found that for
the typical range of current-channel thickness expected in the
corona and used in many MHD simulations \citep{b32a,b30a, b8a,
b25b}, and for a current channel expanding during an upward
perturbation, a critical decay index $n_c$ has similar values for
both the circular and straight current channels in the range 1.1 -
1.3.

\begin{figure*}
\includegraphics[width=167mm]{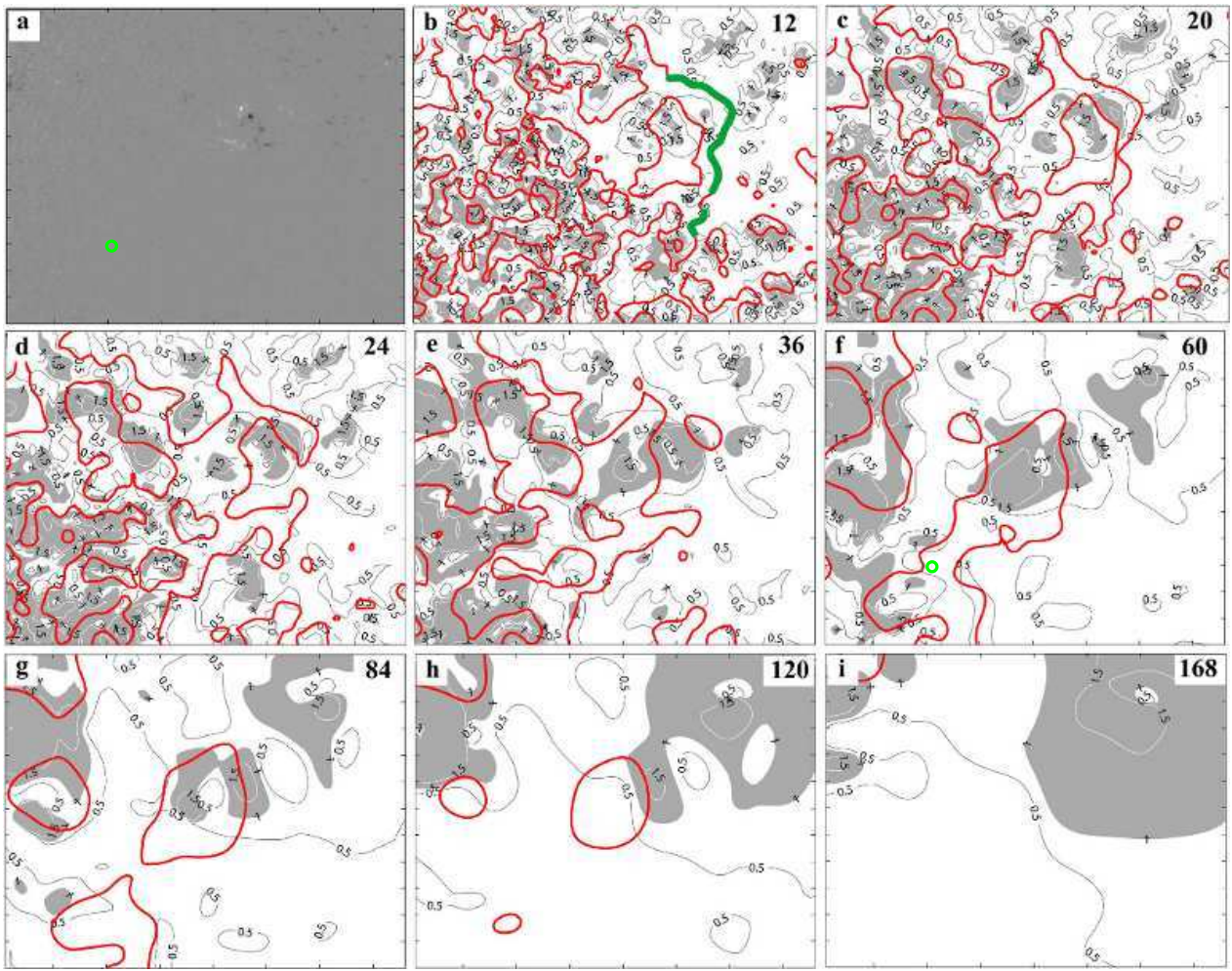}
\caption{Fragment of the {\it SOHO}/MDI magnetogram on 30 April
2010 at 22:27 UT (a) and  distributions of the decay index and
PILs (thick red lines) at different heights indicated at upper
right corners (b) - (i). The thick green line in (b) shows the
section of the PIL line initially occupied by the filament. The
green circle in (b) and (f) shows the final position of the new
filament end region before the filament faded. Shadowed areas show
the regions where $n > 1$. (Courtesy of the {\it SOHO}/MDI
consortium.)}
\end{figure*}

To analyze the equilibrium conditions of the flux rope associated
with the filament we calculated the shape of PILs and the
distribution of the decay index of the potential magnetic field at
different heights in a region surrounding the filament (see
\citealt{b9} for details). Figure 10(a) represents a fragment of
the magnetogram taken by the Michelson Doppler Imager (MDI;
\citealt{b29a}) onboard {\it SOHO} on 30 April 2010 at 22:27 UT,
which was used as a boundary condition for the potential magnetic
field calculations. In Fig. 10, thin lines show isocontours of $n$
= 0.5, 1, 1.5, while thick red lines indicate the positions of
PILs at respective heights. Areas, where $n > 1$, are shadowed. A
PIL is a place where a coronal electric current (a flux rope) can
find the horizontal equilibrium. The thick green line in Fig.
10(b) shows the section of the PIL occupied by the filament as it
is seen in H$\alpha$ line. The filament lays over the PIL
separating the area of predominantly negative network polarity in
the upper right corner of the frame from active region 11064 near
the centre of the frame. The left part of the magnetogram contains
a mixture of faint small-scale opposite polarities producing
numerous closed PIL contours  (Fig. 10(b) - 10(e)).

The height of the prominence in a stable state before the eruption
is 17 Mm (Fig. 5). At this height, the filament is within the area
of stability $n < 1$ (Figures 9(b) - 9(c)). The location of the
filament is stable in the vertical direction even at much greater
heights, but at a height of 20 Mm the PIL occupied by the filament
touches the PIL surrounding the negative polarity of AR 11064 and
reconnects with it. A newly formed PIL (Fig. 10(d)) protrudes far
to the East from the initial filament position. Since the flux
rope can find the horizontal equilibrium only on the PIL, it will
be pushed to the East by the Lorentz force, if it reaches a height
greater than 20 Mm. Moreover, some sections of higher-altitude
PILs are located within unstable (grey) areas. So, the flux rope,
if it arrives upon these sections, will be forced to rise.

PILs associated with small-scale magnetic sources disappear above
a height of 60 Mm. Only a small annular PIL in the centre of the
domain remains at a height of 140 Mm and it also disappears at a
height of 160 Mm.

\section[]{Discussion}

It is widely believed that the most probable initial magnetic
configuration, which accumulates dense prominence plasma and later
hurls a CME into the interplanetary space, is a flux rope
consisted of helical field lines  \citep{b6, b24, b32, b1, b25a,
b22, b39}. An alternative configuration of a CME source region is
a sheared arcade  \citep{b26, b7, b2}, which is converted into a
flux rope structure due to reconnection in the course of the
eruption. Usually, initiation of filament eruptions is associated
with the instability of the flux-rope equilibrium in the vertical
direction. Our prominence was rather low before the eruption, and
it starts to ascend rapidly from a height of about 20 Mm. The maps
of the decay index distribution (Fig. 10) show that the vertical
equilibrium of a flux rope at the location of the filament can be
stable even at several times greater heights. However, the
topology of the magnetic field varies rapidly with height. At a
height of about 20 Mm the PIL, where the flux rope sits, touches
the PIL surrounding the negative polarity of AR 11064 and
reconnects with it. The horizontal equilibrium of the flux rope
can easily be disturbed due to the proximity of another PIL. In
principle, it could find new stable equilibrium over the changed
PIL, but some sections of this PIL are unstable for vertical
displacements, which leads to the filament eruption. This is
unusual initiation of an eruption with combination of initial
horizontal and vertical flux-rope displacements, to the author's
knowledge, never mentioned before.

Initial motion is slow, so it is not so easy to catch the
beginning of the displacements in horizontal and vertical
directions. Taking into account changes in seeing conditions, some
lateral displacement of the filament can be recognized between
22:40 UT and 22:58 UT in MLSO H$\alpha$ images. However, internal
motions within the filament can mask or emulate the real
displacement of the filament as a whole. The same changes can be
found in {\it PROBA2}/SWAP 174  \AA \ images. Small changes in a
height of the prominence are noticeable between 22:36 UT and 22:56
UT in {\it STEREO A} 304 \AA \  images. It is also difficult to
distinguish slow prominence rising from effects related to the
solar rotation and prominence shape changes. Thus, it can be only
stated that both motions start practically simultaneously. We can
compare finite displacements of the filament in horizontal and
vertical directions in the beginning of the event over a period of
time from 23:06 UT to 23:46 UT. Since the contrast of the filament
is low in  the H$\alpha$ images (Fig. 2), we draw in Fig. 11
filament spines with the help of the time sequence (movie 2).  The
horizontal distance is about 150 Mm, while the change in height as
evident from Fig. 3 and Fig. 5  is about 150 Mm too. So, the
deviation of the trajectory of the eruptive filament from the
vertical line is near 45$^\circ$.

\begin{figure}
\includegraphics[width=84mm]{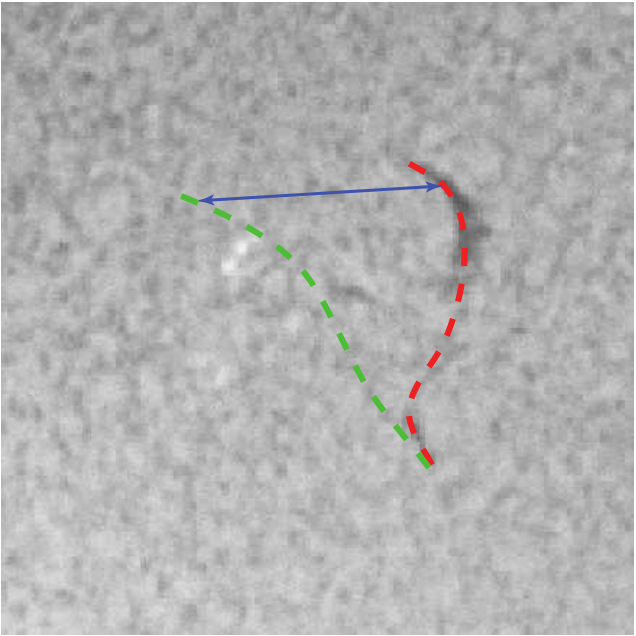}
\caption{Filament spine at 23:06 UT (red line) and 23:46 UT (green
line) superposed on the H$\alpha$ image at 23:06 UT. The blue
arrow shows the horizontal displacement of about 150 Mm. The size
of the frame is .5 $R_\odot$ $\times$ .5 $R_\odot$. (Courtesy of
the Mauna Loa Solar Observatory, operated by the High Altitude
Observatory, as part of the National Center for Atmospheric
Research (NCAR). NCAR is supported by the National Science
Foundation.)}
\end{figure}

The behaviour of the filament endpoints is also unusual. The
northern end of the filament during the eruption suddenly changes
its anchoring to a position about 40$^\circ$ from the starting
position. The process of the filament endpoint change does not
look very similar to whipping-like or zipping-like asymmetric
filament eruptions. The latter would imply the existence of a long
flux rope partly loaded with the filament mass. During the
eruption, internal plasma motions along the flux-rope axis would
reveal the previously invisible far endpoint of the flux rope. In
our case, it is difficult to admit the existence of such a long
flux rope reaching the disk centre in the {\it STEREO B} images.
The PIL, which the filament is associated with, runs to the
North-East at all heights below 40 Mm. There is a high PIL
stretched from the filament position to the South-East (Fig.
10(f)) but we have no manifestations of the presence of the flux
rope there. In principle, we can imagine the existence of an
invisible (no mass load) flux rope with one endpoint anchored at
the bottom-left corner of the studied region, which extends at a
height more than 40 Mm to the upper-right corner where it is lowed
down to a height of 20 Mm and turns to the bottom of the region
providing conditions for the filament formation. Only in this
case,  the observed prominence behaviour could be interpreted as
the asymmetric eruption.

Another possibility is reconnection of the flux-rope field lines
somewhere along the northern half of rising filament loop with
ambient coronal magnetic field lines rooted in a quiet region to
the South-East from the filament. In contrast to examples of the
coronal reconnection, where the filament body disintegrates, the
studied erupting filament loop keeps its shape of a rather thin
loop. Reconnection of an erupting flux rope with ambient coronal
flux has been seen in several numerical simulations of eruptions
\citep{b14a,b14b,b25b}.

We found from comparison of {\it STEREO A} and {\it STEREO B}
images that the observed final height of the originally northern
leg (marked with the blue arrows in Fig. 9) is at least lower than
100 Mm while the height of the prominence top at this time is
about 500 Mm. Thus, the shape of the erupting filament axis is a
loop with the high summit and low ends.

\section[]{Summary and Conclusions}

We studied the filament eruption on 30 April - 1 May 2010, which
shows the reconnection of one filament leg with the region far
away from its initial position. Observations from three viewpoints
were used, namely on-disk H$\alpha$ observations by the Mauna Loa
Solar Observatory, Culgoora Solar Observatory together with EUV
observations by {\it PROBA2} and on-limb observations by {\it
STEREO A} and {\it STEREO B}. At the beginning of the event, the
eruptive prominence looked like a typical one, that is as an
expanding loop with anchored endpoints. Then the top of the loop
folded over showing writhing of the filament axis and deflected to
the South during the ascending. A little later, the endpoints
reshuffled their positions in the limb view. The northern leg of
the erupting prominence loop 'jumps' laterally to the latitude
lower than the latitude of the former southern endpoint. This
behaviour could be interpreted as the asymmetric zipping-like
eruption, although it does not look very likely. Although there is
a PIL stretched at a height above 40 Mm from the former filament
position to the new eastern end point (Fig. 10(f)), no
manifestations of the presence of the flux rope were observed
there and any flare-ribbon-like brightening along the path of this
PIL was absent. Hence, observations do not support the
interpretation of the event as a whipping-like or a zipping-like
asymmetric filament eruption. More probable seems reconnection of
the flux-rope field lines in its northern leg with ambient coronal
magnetic field lines rooted in a quiet region far from the
filament.

The eruption of the prominence was followed by a CME. Due to the
geometrical factor, it was brightest in the field-of-view of the
{\it STEREO A} COR2 coronagraph. At the late stage, the legs of
the CME connecting its core with the Sun show noticeable twisted
structure.

We calculated parameters of coronal potential magnetic field and
found that the eruption is likely to begin with an instability not
in the vertical direction, as typical for eruptive filaments, but
after the violation of the horizontal equilibrium. Observations
show that the trajectory of the eruptive filament deviates from
the vertical line by an angle of about 45$^\circ$. This is unusual
initiation of an eruption with combination of horizontal and
vertical initial flux-rope displacements showing new unexpected
possibility for the start of eruptive events.

\section*{Acknowledgments}

The author acknowledge the Mauna Loa Solar Observatory, Culgoora
Solar Observatory, {\it STEREO}, {\it SOHO}, and {\it PROBA2}
teams for the high-quality data supplied. The author thank the
referee for critical comments and useful suggestions. This work
was supported in part by the Russian Foundation for Basic Research
(grants 12-02-00008, 14-02-92690).

\bsp

\label{lastpage}

\end{document}